\documentclass[onecolumn,showpacs,11pt]{revtex4-2}
\usepackage{mathrsfs}
\usepackage{graphicx}
\usepackage{bm}
\usepackage{float}
\usepackage{fancyhdr}
\usepackage{longtable}
\usepackage{hyperref}
\usepackage{subfigure}
\usepackage{epsfig}
\usepackage{color,soul}
\usepackage{menukeys}
\usepackage{xcolor}
\usepackage{feynman}
\usepackage{amssymb}
\usepackage{braket}
\usepackage{graphicx}
\usepackage{dcolumn}
\usepackage[mathscr]{eucal}
\usepackage{amsmath}
\usepackage{tabularx}
\usepackage{booktabs}
\usepackage{slashed}
\usepackage{mathtools,slashed}
\usepackage{axodraw2}

\allowdisplaybreaks

\makeatletter
\def\pslashed#1{\expandafter\ifx\csname psla@\string#1\endcsname\relax
{\mathpalette{\sla@/00}{\phantom{#1}}}\else
\csname psla@\string#1\endcsname\fi}
\def\declarepslashed#1#2#3#4#5{\expandafter\def\csname psla@\string#5\endcsname{#1{\mathpalette{\sla@{#2}{#3}{#4}}{\phantom{#5}}}}}
\makeatother
\declarepslashed{}{/}{.08}{0}{D}
\newcommand{\hs}{\hspace*{0.5cm}}

\newcommand{\be}{\begin{equation}}
\newcommand{\ee}{\end{equation}}
\newcommand{\bea}{\begin{eqnarray}}
\newcommand{\eea}{\end{eqnarray}}
\newcommand{\ben}{\begin{enumerate}}
\newcommand{\een}{\end{enumerate}}
\newcommand{\bde}{\begin{widetext}}
\newcommand{\ede}{\end{widetext}}
\newcommand{\nn}{\nonumber}
\newcommand{\crn}{\nonumber \\}

\newcommand{\al}{\alpha}

\newcommand{\fr}{\frac}
\newcommand{\bc}{\begin{center}}
\newcommand{\ec}{\end{center}}
\newcommand{\Ga}{\Gamma}

\newcommand{\La}{\Lambda}

\begin{document}

\title{Embedding light dark matter and small neutrino mass\\ in the flipped standard model}
\author{D. T. Huong}
\email{dthuong@iop.vast.vn}
\affiliation{Institute of Physics, Vietnam Academy of Science and Technology, 10 Dao Tan, Giang Vo, Hanoi 100000, Vietnam}
\author{Phung Van Dong}
\email{dong.phungvan@phenikaa-uni.edu.vn}
\affiliation{Phenikaa Institute for Advanced Study, Phenikaa University, Nguyen Trac, Duong Noi, Hanoi 100000, Vietnam}
\author{A. E. C\'{a}rcamo Hern\'{a}ndez$^{a,b,c}$}
\email{antonio.carcamo@usm.cl}
\affiliation{$^{{a}}$Universidad T\'{e}cnica Federico Santa Mar\'{\i}a, Casilla 110-V, Valpara%
\'{\i}so, Chile\\
$^{{b}}$Centro Cient\'ifico-Tecnol\'ogico de Valpara\'iso, Casilla 110-V, Valpara%
\'{\i}so, Chile\\
$^{{c}}$Millennium Institute for Subatomic physics at high energy frontier - SAPHIR,
Fernandez Concha 700, Santiago, Chile.}
\date{\today}

\begin{abstract}
We revisit the flipped standard model where a $U(1)_N$ gauge group is added, determining a dark charge through the weak isospin such as $D=T_3+N$, analogous to the electric charge and hypercharge relation. We find 
that neutrino masses are appropriately generated by a radiative inverse seesaw mechanism mediated by dark fields. Dark matter candidate is a naturally light fermion with the mass radiatively induced at the keV scale. The residual $Z_2$ parity arising from $U(1)_N$ symmetry breaking both stabilizes the dark matter candidate and prevents its potential mixing with neutrinos. Such residual $Z_2$ parity also guarantees the radiative nature of the inverse seesaw mechanism responsible for light active neutrino mass generation. It is noted that the keV dark matter may be thermally produced in the early Universe as decoupled but being still relativistic and typically overpopulated due to $U(1)_N$ portal interactions. To achieve the correct abundance, the excessive thermal production is counterbalanced by sufficient late-time entropy generation from the decay of long-lived particles. The parameter space under consideration can simultaneously accommodate the observational data from cosmic inflation and keV dark matter.
\end{abstract}

\maketitle

\section{Introduction} 
The standard model (SM) of electroweak and strong interactions has been greatly successful in describing the observed phenomena with high accuracy. In spite of this success, there remain several issues suggesting that the SM must be extended. The first of which is existence of tiny neutrino masses, which lead to neutrino oscillations, as confirmed by experiment \cite{Kajita:2016cak,McDonald:2016ixn}. The second of which is presence of dark matter (DM) abundance, which makes up most of the mass of galaxies and galactic clusters, as observed in the Universe \cite{Bertone2005279,Arcadi:2017kky}. Both the issues do not find any explanation within the conventional framework of the SM. 

The seesaw mechanism imposes heavy right-handed neutrinos coupled to left-handed neutrinos via the usual Higgs field, hence producing small neutrino masses as suppressed by the right-handed neutrino mass scale \cite{Minkowski:1977sc,GellMann:1980vs,Yanagida:1979as,Mohapatra:1979ia,Schechter:1980gr}. However, this seesaw scale has to be fourteen orders of magnitude compared to the weak scale, given that the Dirac masses which couple left-handed neutrinos to right-handed neutrinos are at the weak scale as usual fermions, making the mechanism untestable at experiment. Besides, the seesaw mechanism in itself cannot explain the existence of DM. This motivates modifications of the seesaw mechanism with included discrete (or gauge) symmetry and extended particle spectrum, which implement a radiative low-scale seesaw instead of the tree-level seesaw responsible for neutrino mass generation and DM stability \cite{Ma:2006km,Okada:2010wd,Montero:2011jk,Okada:2012sg,Basak:2013cga,Sanchez-Vega:2014rka,Okada:2016tci,Rodejohann:2015lca,Okada:2019sbb,Okada:2020cvq,Dasgupta:2019rmf,Biswas:2019ygr,Gehrlein:2019iwl,Han:2020oet,Choudhury:2020cpm,Mahapatra:2020dgk,Leite:2020wjl,Okada:2020cue,Motz:2020ddk}. Alternative extensions of the SM rely on a global symmetry whose spontaneous breaking if occurs yields a preserved discrete symmetry guaranteeing the radiative seesaw nature for neutrino mass as well as DM stability \cite{CarcamoHernandez:2023atk,Abada:2022dvm,Abada:2023zbb,Bonilla:2023wok,Bonilla:2023egs,Benitez-Irarrazabal:2025atb,Abada:2025ozu}. 

It is well established that according to the SM extension for neutrino mass, e.g. $U(1)_{B-L}$ \cite{Davidson:1978pm,Mohapatra:1980qe, Marshak:1979fm}, $U(1)_{L_i-L_j}$ \cite{Foot:1990mn,Foot:1994vd,He:1991qd}, and \cite{Holdom:1985ag,Appelquist:2002mw}, DM candidates also emerge in variety of ways, sometimes introducing extra particles or discrete symmetries by {\it ad hoc}, such that the symmetry assignment ensuring DM stability is fully elective. This motivates studies about the original nature of discrete symmetry that makes DM stable, for instance \cite{Dong:2018aak,PhysRevD.103.095016,PhysRevD.79.041701,PhysRevD.81.051702,VanDong:2020cjf,VanLoi:2020kdk,VanLoi:2021dra}. In the gauge approach, the breaking of the new gauge symmetry determines both neutrino mass and DM stability. As a matter of fact, a preserved discrete symmetry arising as the residual new gauge symmetry dictates DM stability. The dynamics of the new gauge symmetry governs DM observables. However, the DM mass is often given at the new symmetry breaking scale, which is beyond the weak scale. In this paper, we propose how a new gauge extension of the SM can naturally provide a light DM candidate, by contrast, besides providing appropriate small neutrino masses, which interestingly come from an inverse seesaw \cite{mohapatra:1986aw,Mohapatra:1986bd,bernabeu:1987gr}.

Among the many extensions of the Standard Model gauge symmetry, the existence of a new abelian gauge group factor is particularly intriguing. From the topdown approach, the unified gauge theories hint such a symmetry. For instance, left-right symmetry, Pati-Salam unification, $SO(10)$, trinification, and $E_6$ indicate to a $U(1)_{B-L}$ group in which $B-L$ is the baryon minus lepton number, which is always preserved by low-energy processes. The unification of fermion families---as inspired by lepton and quark mixings---by a discrete or gauge symmetry implies a conserved abelian charge, such as $L_i-L_j$ or $B_i-B_j$, where $L_i$ $(B_i)$ is the lepton (baryon) number of flavor $i$. Now that a dark charge necessarily exists in order to stabilize DM and set its observables. The dark charge might arise from a unified theory, imprinted at low energy as residual $U(1)$ factor \cite{VanDong:2025kxl}. As in the dark photon theory \cite{Holdom:1985ag,Appelquist:2002mw}, the most new charges reduced from the topdown or put by hand are abelian, which is a result of the proper embedding of the standard model gauge symmetry in extended symmetries. 

An intriguing alternative approach is given upon the nature of the dequantization of the electric charge ($Q$) in the SM, which reveals a non-abelian dark charge. It postulates that a dark charge ($D$) may exist as a dequantized version of the electric charge. It means that a non-abelian dark charge $D$ exists which neither commutes nor closes algebraically with weak isospin, i.e. $[D,T_1\pm i T_2]=\pm (T_1\pm i T_2)$, analogous to the electric charge \cite{VanDong:2020cjf,VanLoi:2020kdk}. The algebraic closure of the electric charge with the weak isospin demands that the (abelian) hypercharge exists so that $Q=T_3+Y$, which is well-established in the standard model. Similarly, the algebraic closure of the dark charge with the weak isospin yields a new abelian hypercharge (called hyperdark) as such $D=T_3+N$. The resultant theory is given by the gauge symmetry $SU(3)_C\otimes SU(2)_L \otimes U(1)_Y\otimes U(1)_N$, so-called flipped standard model. In this work, we revisit this theory, showing that neutrinos can obtain an appropriate mass from a radiative inverse seesaw, by contrast. Furthermore, among the dark fields contributing to the neutrino mass, a light DM is naturally emerged with a mass radiatively generated at keV. This is opposite to the original proposal in \cite{VanDong:2020cjf,VanLoi:2020kdk}, in which the neutrino mass was produced by a canonical seesaw, whereas DM candidates were separately, {\it ad hoc} included. 

The rest of this work is organized as follows. In Sec. \ref{model}, we rebuild the flipped standard model with stressing that the novel particle assignment gives rise to the interplay between neutrino mass generation and DM existance. In Sec. \ref{neudm}, we present the results of neutrino mass generation and DM stability, in which the relevant DM observables are examined. In Sec. \ref{concl}, we summarize our results and conclude this work.                 

\section{\label{model} Revisiting the flipped standard model} 

\begin{table}[h]
\centering 
\begin{tabular}{rccccccccccc}
\hline\hline
Multiplet & $Q_{aL}$ & $u_{aR}$ & $d_{aR}$ & $L_{aL}$ & $e_{aR}$ & $\nu _{aR}$ & $N_{1} $ & $N_{2}$ & $S_{1}$ & $S_{2}$ & $S_{3}$ \\ \hline
$SU(3)_{C}$ & $3$ & $3$ & $3$ & $1$ & $1$ & $1$ & $1$ & $1$ & $1$ & $1$ & $1$ \\ 
$SU(2)_{L}$ & $2$ & $1$ & $1$ & $2$ & $1$ & $1$ & $1$ & $1$ & $1$ & $1$ & $1$ \\ 
$U(1)_{Y}$ & $\frac{1}{6}$ & $\frac{2}{3}$ & $-\frac{1}{3}$ & $-\frac{1}{2}$ & $-1$ & $0$ & $0$ & $0$ & $0$ & $0$ & $0$ \\ 
$U(1)_{N}$ & $\frac{1}{6}-\frac{5\delta }{3}$ & $\frac{2}{3}-\frac{5\delta }{3}$ & $-\frac{1}{3}-\frac{5\delta }{3}$ & $5\delta -\frac{1}{2}$ & $5\delta -1$ & $5\delta$ & $-7\delta$ & $-9\delta$ & $2\delta$ & $4\delta$ & $10\delta $ \\ \hline\hline
\end{tabular}%
\caption{Fermion representations according to $SU\left( 3\right) _{C}\otimes SU\left(3\right) _{L}\otimes U\left( 1\right) _{Y}\otimes U(1)_N$. }
\label{fermions}
\end{table}

\begin{table}[h]
\centering 
\begin{tabular}{rcccccc}
\hline\hline
Multiplet & $\phi$ & $\eta _{1}$ & $\eta _{2}$ & $\varphi _{1}$ & $\varphi _{2}$ & $\varphi_{3}$  \\ \hline
$SU(3)_{C}$ & $1$ & $1$ & $1$ & $1$ & $1$ & $1$ \\ 
$SU(2)_{L}$ & $2$ & $1$ & $1$ & $1$ & $1$ & $1$ \\ 
$U(1)_{Y}$ & $\frac{1}{2}$ & $0$ & $0$ & $0$ & $0$ & $0$ \\ 
$U(1)_{N}$ & $\frac{1}{2}$ & $2\delta$ & $4\delta$  & $5\delta$ & $-\delta$ & $-9\delta$ \\ \hline\hline
\end{tabular}
\caption{Scalar representations according to $SU\left( 3\right) _{C}\otimes SU\left(3\right) _{L}\otimes U\left( 1\right) _{X}\otimes U(1)_{N}$.}
\label{scalars}
\end{table}

We consider the SM extension with the $SU(3)_C\otimes SU(2)_{L}\otimes U(1)_{Y}\otimes U(1)_{N}$ gauge symmetry, where the first two groups correspond to the color charge and the weak isospin, respectively. The hypercharge $Y$ and the hyperdark $N$ determine the electric charge $Q$ and the dark charge $D$, such as $Y=Q-T_{3}$ and $N=D-T_{3}$, analogous to \cite{VanLoi:2020kdk,VanLoi:2021dra}, respectively. It is noted that anomaly cancellation requires the existence of 
the right handed neutrinos $\nu_{aR}$, $N_n$ ($n=1,2$) and $S_a$  ($a=1,2,3$), with the non trivial $U(1)_{N}$ charge assignments specified in Table \ref{fermions}.  
It is assumed that these right-handed neutrinos cannot receive a Majorana mass through dark charge breaking, which suppresses the possibility of neutrino mass generation by canonical seesaw. In the present theory, we introduce two fermion singlets $N_{1,2}$ as coupled to the right-handed neutrinos in order to produce appropriate neutrino masses via inverse seesaw. Here, the small Majorana masses of $N_{1,2}$ in the inverse seesaw are radiatively generated by two of fermion singlets $S_{1,2,3}$, which are necessarily presented to cancel all anomalies, say $[\text{Gravity}]^{2}U(1)_{N}$ and $[\text{U(1)}]_{N}^{3}$. 
That said, the full fermion representations of the model under the $SU(3)_{C}\otimes SU(2)_{L}\otimes U(1)_{Y}\otimes U(1)_{N}$ gauge symmetry are collected in Table \ref{fermions}. The scalar sector beside the standard model Higgs field $\phi$ includes scalar singlets $\eta_{1,2}$ and $\varphi_{1,2,3}$ necessarily coupled to the mentioned couplings of fermions (not to $\nu_R\nu_R$). The quantum numbers of scalars under the gauge symmetry are presented in Table \ref{scalars}. Explicitly, the fields $\eta_1$ and $\eta_2$ are required to generate Dirac neutrino masses. Additionally, the presence of $\eta_2$ provides tree-level masses to $S_1$ dark fermions. The dark fermions $S_2$ and $S_3$ acquire masses via loop corrections due to virtual exchanges of dark scalars $\varphi_1$, $\varphi_2$, $\varphi_3$ and neutral fermions $N_1$, $N_2$, $S_1$, and $S_2$, as shown in Feynman diagrams in Figure \ref{DMmass}.

As it will be shown in detail below, the scalar fields $\phi$ and $\eta_{1,2}$ will be neutral under the residual dark parity $D_P$, whereas the remaining scalars will be non trivial dark parity states. Consequently, the scalar fields $\phi$ and $\eta_{1,2}$ will acquire non vanishing vacuum expectation values (VEVs), such as $\langle \phi\rangle = (0,v/\sqrt{2})^T$ and $\langle \eta_{1,2}\rangle=\La_{1,2}/\sqrt{2}$, whereas $\varphi_{1,2,3}$ do not, i.e. $\langle \varphi_{1,2,3}\rangle =0$. We now find a dark parity that emerges naturally as a residual symmetry resulting from the spontaneous breaking of the gauge symmetry. Indeed, the process of spontaneous symmetry breaking proceeds as follows 
\begin{center}
	\begin{tabular}{c}
		$SU(3)_C\otimes SU(2)_L\otimes U(1)_Y\otimes U(1)_{N}$ \\ 
		$\downarrow \Lambda_{1,2}$ \\ 
		$SU(3)_C\otimes SU(2)_L\otimes U(1)_Y\otimes  N_P$ \\ 
		$\downarrow  v $ \\ 
		$SU(3)_C\otimes U(1)_Q\otimes D_P$
	\end{tabular}
\end{center}
Each stage gives rise to a distinct residual discrete symmetry, such as $N_P$ and $D_P$. They emerge sequentially at different energy scales throughout the symmetry breaking process. The initial breaking step involves the spontaneous breaking of the $U(1)_N$ gauge symmetry, which is derived by the VEVs of $\eta_{1,2}$. This leads to the residual discrete symmetry $N_P$, such as $N_P=(-1)^{\frac{kN}{ \delta}}$, where $k$ is integer. The subsequent breaking step breaks the electroweak symmetry as well as $N_P$, derived by the usual Higgs field. This leads to the final residual discrete symmetry, $D_P= (-1)^{\frac{kD}{\delta}}$, where $D=T_3+N$. The form of $D_P$ exists as a mixture of the weak isospin and the hyperdark, implying that the dark charge is nonabelian, as mentioned. Notice that the half-integer $U(1)_N$ charge of the Higgs doublet implies that the weak vacuum conserves dark charge.

We assume that $\delta = p/m$, where $m$ and $p$ are integers. After the breaking of the gauge symmetry, the residual symmetry is a discrete cyclic group $Z_k$. This implies that all matter fields must satisfy
$k\frac{\pi D}{\delta} = n_i 2\pi$ with $D=  D_u, D_d,D_e, D_{H^+}, D_\nu$ corespond to $i=1,...,5$ and $n_i$'s are an integer number. This condition leads to:
\bea
&& -\frac{5}{3}+\frac{2m}{3p}=\frac{2 n_1}{k}, \hs -\frac{m+5p}{3p}=\frac{2n_2}{k}, \crn 
&& 5-\frac{m}{p} =\frac{2n_3}{k}, \hs \frac{m}{p} =\frac{2n_4}{k},
\hs \frac{2 n_5}{k}= 5.
\eea
Solving the system yields:
\bea
&& m=\frac{2(n_1-n_2)p}{k}, \hs  k=-\frac{2}{5} \left(n_1+2n_2 \right), \crn
&& n_3=-2n_1-n_2, \hs n_4 =n_1-n_2, \hs  n_5 =\frac{5}{2}k.
\eea
This implies that $k$ is even. Hence, the residual symmetry $D_P$ is determined to be an even cyclic group. The smallest group of which is with $k=2$, obtained by a dark parity
$P=\left(-1\right)^{\frac{m}{p}D}=\left(-1\right)^{\frac{1}{2}\left(5+3n_1\right)D}$. We multiply the dark parity $P$ with the spin parity $S = (-1)^{2s}$, which is always conserved, to redefine the residual symmetry conveniently $D_P = Z_2 = \{1, PS\}$. In short, the theory is governed by a dark parity, such that 
\begin{equation}
	D_P \equiv PS= (-1)^{\frac{1}{2}\left(5+3n_1\right)D+2s},
\end{equation} where $D$ and $s$ are dark charge and spin, respectively.   

Remarks are in order:
\begin{itemize}
	\item If $n_1=4m+1$, then all SM particles and new particles $\nu_{aR}, N_{1,2}, \eta_{1,2}$ are even under $Z_2$, whereas the particles $S_{1,2,3}, \varphi_{1,2,3}$ are odd.
	\item If $n_1=4m+3$, then up-type components of doublets and singlets $\nu_{aR}, N_{1,2}, \eta_{1,2}$ are even under $Z_2$, whereas down-type components of doublets and singlets $S_{1,2,3}, \varphi_{1,2,3}$ are odd.
	\item If $n_1=4m$ or $4m+2$, then some physical fields cannot be any eigenstate of the $Z_2$ symmetry defined by the generator  $D_P = (-1)^{\frac{1}{2}\left(5+3n_1\right)D+2s}$. Hence, the residual symmetry must be larger than the present $Z_2$.
\end{itemize}
So among the minimal choices of the residual symmetry, we will take the case $n_1=4m+1$ into account. Interestingly, this residual symmetry plays the crucial roles, ensuring both the stability of dark matter candidates and the radiative origin of the inverse seesaw responsible for small neutrino masses. It is noteworthy that the scalar singlets $\varphi_1$, $\varphi_2$ and $\varphi_3$ do not acquire any VEV as they carry non-trivial dark charges protected by the dark parity $D_P$. This symmetry is also essential to forbid tree-level mass terms, which otherwise mix the even fields $N_{1,2}, \nu_{aR}$ with the odd fields $S_{1,2,3}$. Such odd fermions would generate effective Majorana mass terms like $\bar{N}_{1}^{c}N_1$ and $\bar{N}_{2}^{c}N_2$ leading to the realization of the radiative inverse seesaw.
  
\section{\label{neudm} Neutrino mass generation and dark matter stability}

From the particle representation content under the previously described relevant symmetries, we can build the following Yukawa interactions for charged fermions: 
\begin{equation}
\mathcal{L}_{Yukawa}^{c.f.}=h_{ab}^{e}\bar{L}_{aL}\phi e_{bR}+h_{ab}^{d}\bar{Q}_{aL}\phi d_{bR}+h_{ab}^{u}\bar{Q}_{aL}\tilde{\phi}u_{bR}+H.c.,
\end{equation} while the Yukawa interactions for neutral fermions are given by
\begin{eqnarray}
\mathcal{L}_{Yukawa}^{n.f.} &=&h_{ab}^{\nu }\bar{L}_{aL}\tilde{\phi}\nu_{bR}+f_{a1}^{\nu }\bar{\nu}_{aR}^{c}\eta _{1}N_{1}+f_{a2}^{\nu }\bar{\nu}_{aR}^{c}\eta _{2}N_{2}+f_{1}^{N}\bar{S}_{1}^{c}N_{1}\varphi _{1}+f_{2}^{N}\bar{S}_{2}^{c}N_{2}\varphi _{1}\notag \\
&&+\frac{1}{2}f_{1}^{S}\bar{S}_{1}^{c}\eta _{2}^{\ast }S_{1}+\frac{1}{2}f_{2a}^{S}\bar{S}_{2}^{c}\varphi _{3}\nu_{aR}+\frac{1}{2}f_{3}^{S}\bar{S}_{3}^{c}\varphi _{2}N_{2}+H.c.
\label{newDM}
\end{eqnarray}%

After the breaking of the $SU(2)_{L}\otimes U(1)_{Y}\otimes U(1)_{N}$ gauge symmetry, the charged fermions, such as quarks and charged leptons, gain an appropriate mass from the charged fermion Yukawa interactions, similar to the Standard Model. On the other hand, the neutral fermion Yukawa interactions yield the following mass terms for visible neutrinos: 
\begin{equation}
\mathcal{L}_{mass}^{neutrino}=-\bar{\Psi}\mathcal{M}_{\nu }\Psi +H.c =-\left( \begin{array}{ccc}
\bar{\nu}_{L} & \bar{\nu}_{R}^{c} & \bar{N}^c \end{array} \right) \left( \begin{array}{ccc}
0 & M_{D} & 0 \\ 
M_{D}^{T} & 0 & M_{N} \\ 
0 & M_{N}^{T} & \mu \end{array} \right) \left( \begin{array}{c}
\nu _{L} \\ \nu _{R} \\ N \end{array} \right)+H.c,
\end{equation}%
where the entries of Dirac $M_{D}$ and Majorana $M_{N}$ submatrices arise at tree level, given by
\begin{equation}
\left( M_{D}\right) _{ab}=-h_{ab}^{\nu }\frac{v}{\sqrt{2}},\hspace{0.5cm}\left( M_{N}\right) _{an}=-f_{an}^{\nu }\frac{\Lambda _{n}}{\sqrt{2}},\hspace{0.5cm}a,b=1,2,3,\hspace{0.5cm}n=1,2.
\end{equation} By contrast, the Majorana submatrix $\mu $, which is $2\times 2$, vanishes at tree level, but it may be generated at one (or two) loop level from the Feynman diagrams shown in Fig. (\ref{mu}). See Appendix~\ref{app1} for estimating the lepton number violating Majorana mass submatrix $\mu$. 
Since $\mu_{nm}<<\left( M_{D}\right) _{ab}<<\left( M_{N}\right) _{an}$ ($n,m=1,2$ and $a,b=1,2,3$) the tiny masses for active neutrinos are produced by a radiative inverse seesaw mechanism. Consequently, the mass matrices of the active and heavy neutrinos are given by
\begin{eqnarray}
M_{\text{active}}&=& M_D \left( M_N \mu^{-1} M_N^{T}\right)^{-1} M_D^T, \\
M_{N^-} &=&-\frac{1}{2}\left(M_N^T M_N \right)^{\frac{1}{2}} +\frac{1}{2}\mu,\\
M_{N^+}&=&\frac{1}{2}\left(M_N M_N^T \right)^{\frac{1}{2}} +\frac{1}{2}\frac{M_N \mu M^T_N}{ M_N M^T_N},
\end{eqnarray}
where $M_{\text{active}}$ is the active neutrino mass matrix, whereas $M_{N^-}$ and $M_{N^+}$ are the exotic neutrino mass matrices. It is noted that $M_{N^{-}}$ is a $2 \times 2$ matrix, while $M_{N^{+}}$ is a $3 \times 3$ matrix. It is stressed that the physical spectrum of the visible neutrino sector contains three active neutrinos and five heavy exotic neutrinos, from which two of them arise from the $M_{N^-}$ matrix, whereas the remaining three originate from $M_{N^+}$. Let us note that the decay of the lightest sterile neutrino in the early Universe can give rise to the measured baryon asymmetry.

\begin{figure}[h]
	\begin{center}
			\includegraphics[scale=0.54]{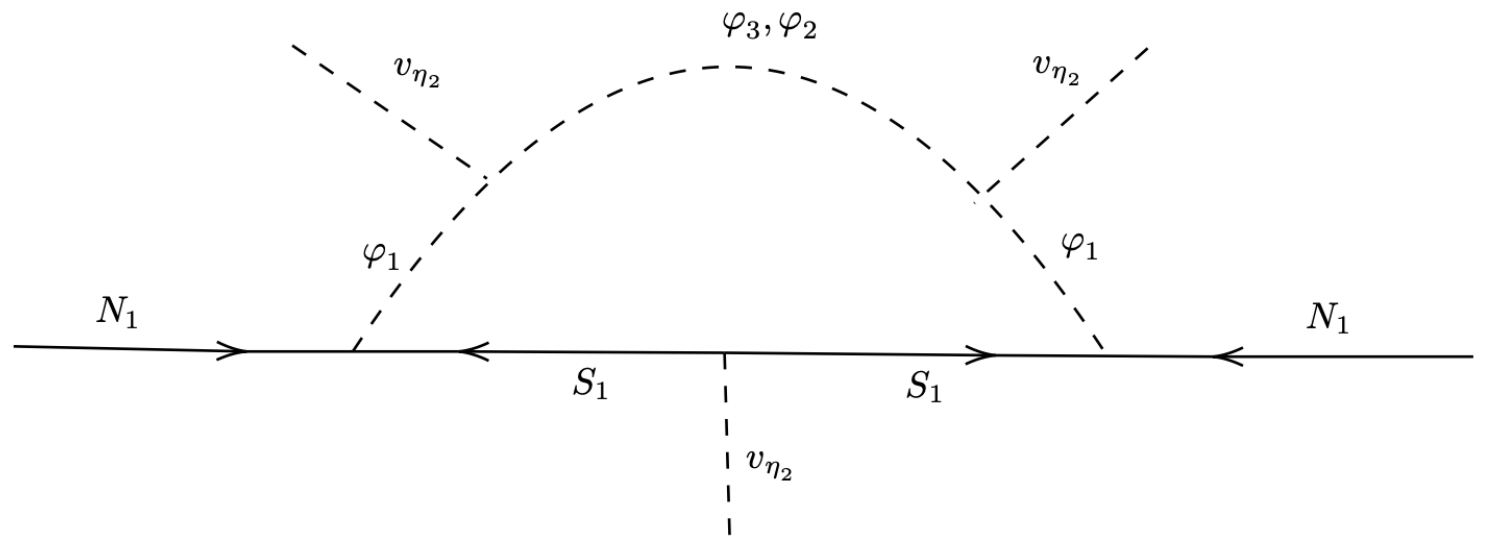} \\
			\includegraphics[scale=0.5]{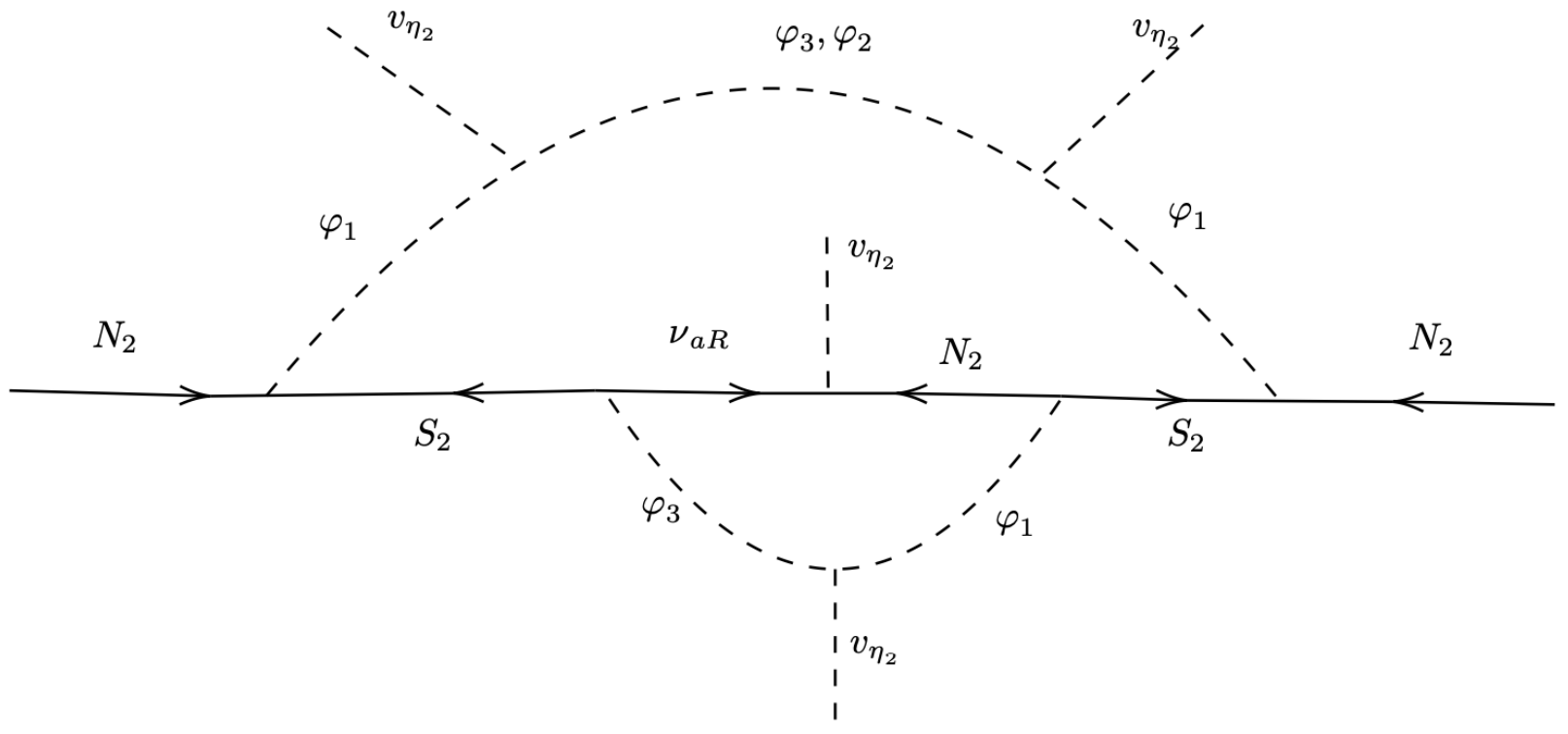}  \\
			\includegraphics[scale=0.5]{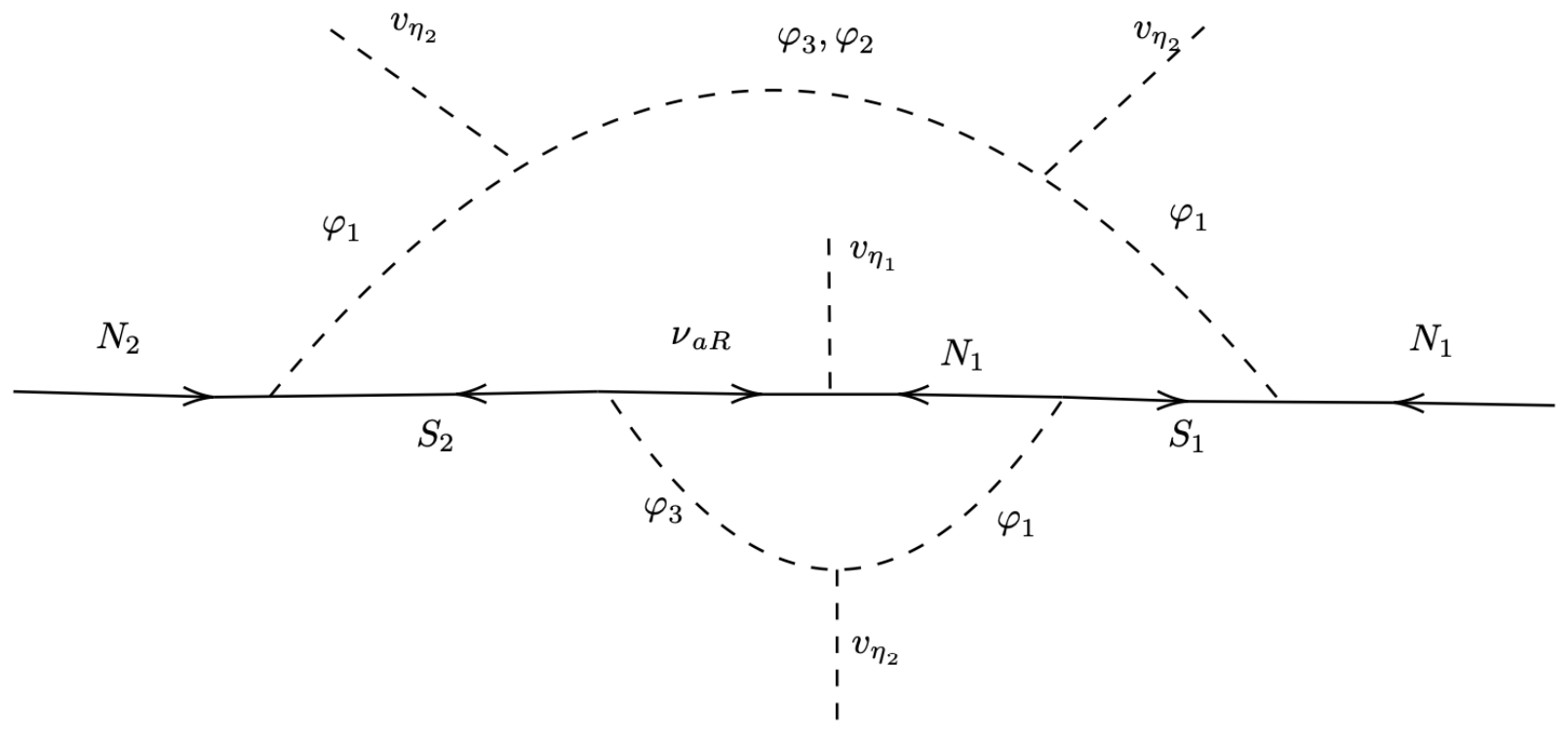} 
	\end{center}
	\caption{Feynman diagrams contributing to the submatrix $\mu$.}
	\label{mu}
\end{figure}

Last, but not least, the neutral fermion Yukawa interactions also provide a mass matrix responsible for dark neutral leptons, obtained as
\begin{equation}
	\mathcal{L}_{m}^{DM}=- \frac{1}{2}\left( \begin{array}{ccc}
		\bar{S}_{1}^c & \bar{S}_{2}^{c} & \bar{S}_3^c \end{array} \right)\left( \begin{array}{ccc} m_{S_1S_1}^{\text{tree}} &m_{S_1S_2}^{\text{one-loop}} &m_{S_1S_3}^{\text{three-loop}} \\ m_{S_2S_1}^{\text{one-loop}}& m_{S_2S_2}^{\text{one-loop}} & m_{S_{2}S_3}^{\text{three-loop}} \\ m_{S_3S_1}^{\text{three-loop}}& m_{S_{3}S_2}^{\text{three-loop}}& m_{S_{3}S_3}^{\text{three-loop}} \end{array} \right)\left( \begin{array}{c} S_1 \\ S_2 \\ S_3 \end{array} \right),
\label{massDM}\end{equation}
where only the (1,1) element is generated at tree level, while the remaining entries vanish at this order. The radiative matrix elements can be computed by evaluating the relevant Feynman diagrams, as shown in Fig.~\ref{DMmass}. Details of the calculation are provided in Appendix~\ref{app2}. The mass matrix in Eq.~\ref{massDM} then yields three massive eigenstates:
$S_1^\prime$ (a heavy particle), 
$S_2^\prime$ (a light particle), 
and $S_3^\prime$ (a very light field), 
which are related to the flavor states by
\[
\left( \begin{array}{c} S_1 \\ S_2 \\ S_3 \end{array} \right)^T = U^S \left( \begin{array}{c} S_1^\prime \\ S_2^\prime \\ S_3^\prime \end{array} \right)^T,
\]
where $U^S$ is a $3 \times 3$ mixing matrix with very suppressed off-diagonal elements. Due to the conservation of the dark parity, the lightest of these states, $S_3^\prime$, stands as a fermionic dark matter candidate. Its mass originates from a three-loop radiative seesaw mechanism, and can acquire values at the keV scale. This can be directly seen by considering the simplifying benchmark scenario $f_{1}^N<<f_{2}^N\simeq f_{2a}^S<<f_{3}^S$, ($a=1,2,3$), where as discussed in Appendix \ref{app2}, the mass for the dark neutral lepton $S_3^\prime$ can be estimated as follows:
\begin{equation}
m_{S_3^\prime}\simeq\frac{\left( f_{2}^{N}\right) ^{3}\left(
	f_{3}^{S}\right) ^{2}}{\left( 16\pi ^{2}\right) ^{3}}\sum_{a=1}^{3}f_{2a}^{S}\left(
M_{N}\right) _{a2}.
\end{equation}
Taking $f_{2}^N\simeq f_{2a}^S\sim \mathcal{O}(0.1)$ ($a=1,2,3$), $f_{3}^S\sim f_{an}^{\nu }\sim\mathcal{O}(1)$ ($a=1,2,3$, $n=1,2$), $\Lambda_1\simeq\Lambda_2\simeq\Lambda\sim\mathcal{O}(10)$ TeV, we find $m_{S_3^\prime}\sim 10^{-10}\Lambda\sim\mathcal{O}(1)$ keV, which shows that our model can naturally accommodate a keV fermionic dark matter candidate.
\begin{figure}[h]
	\begin{center}
			\includegraphics[scale=0.13]{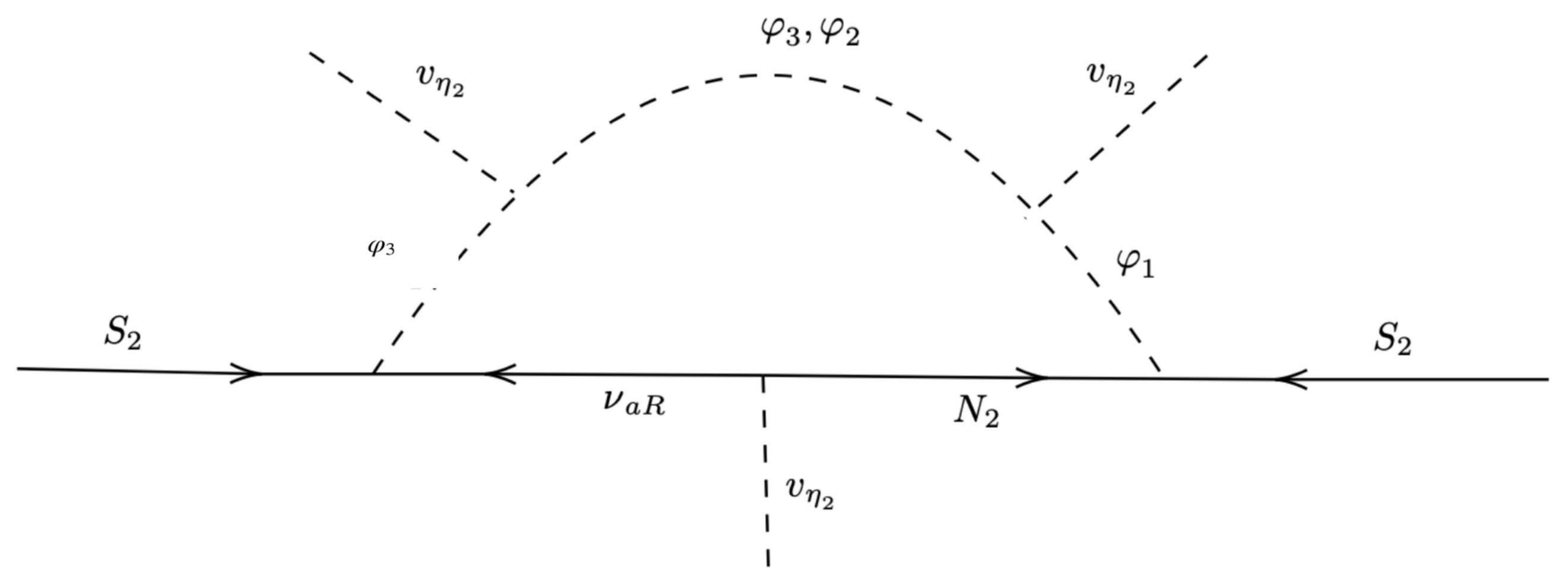} 
			\includegraphics[scale=0.35]{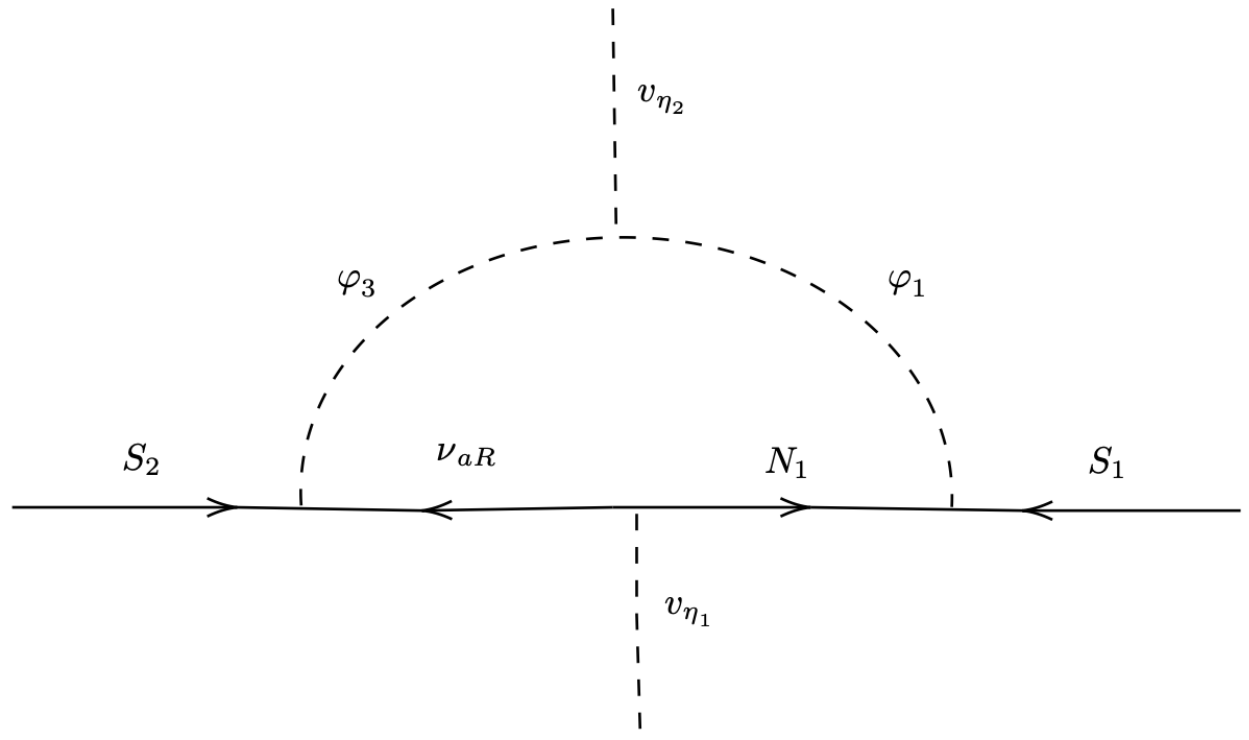}  		   
			\includegraphics[scale=0.28]{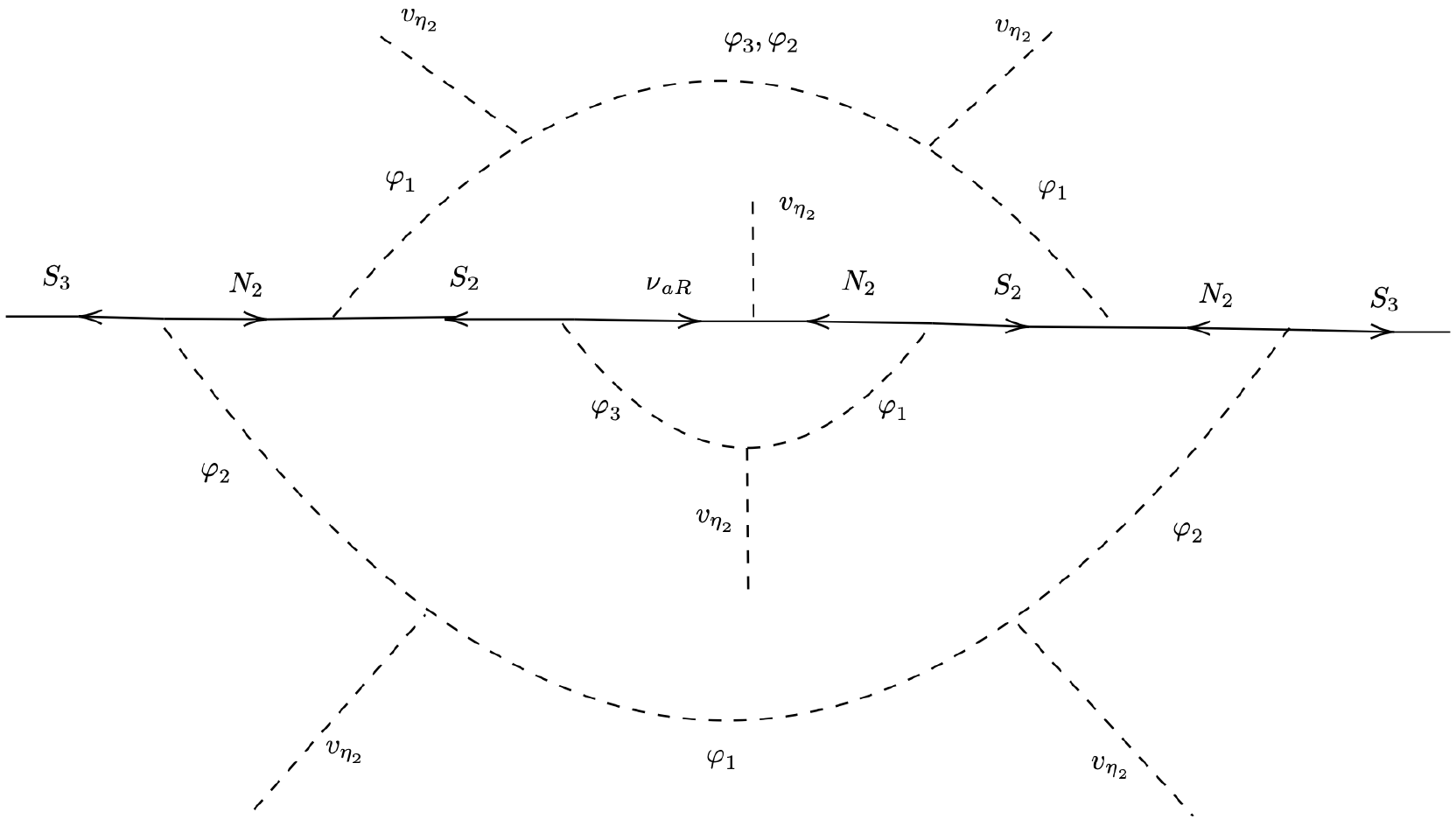}
			\includegraphics[scale=0.24]{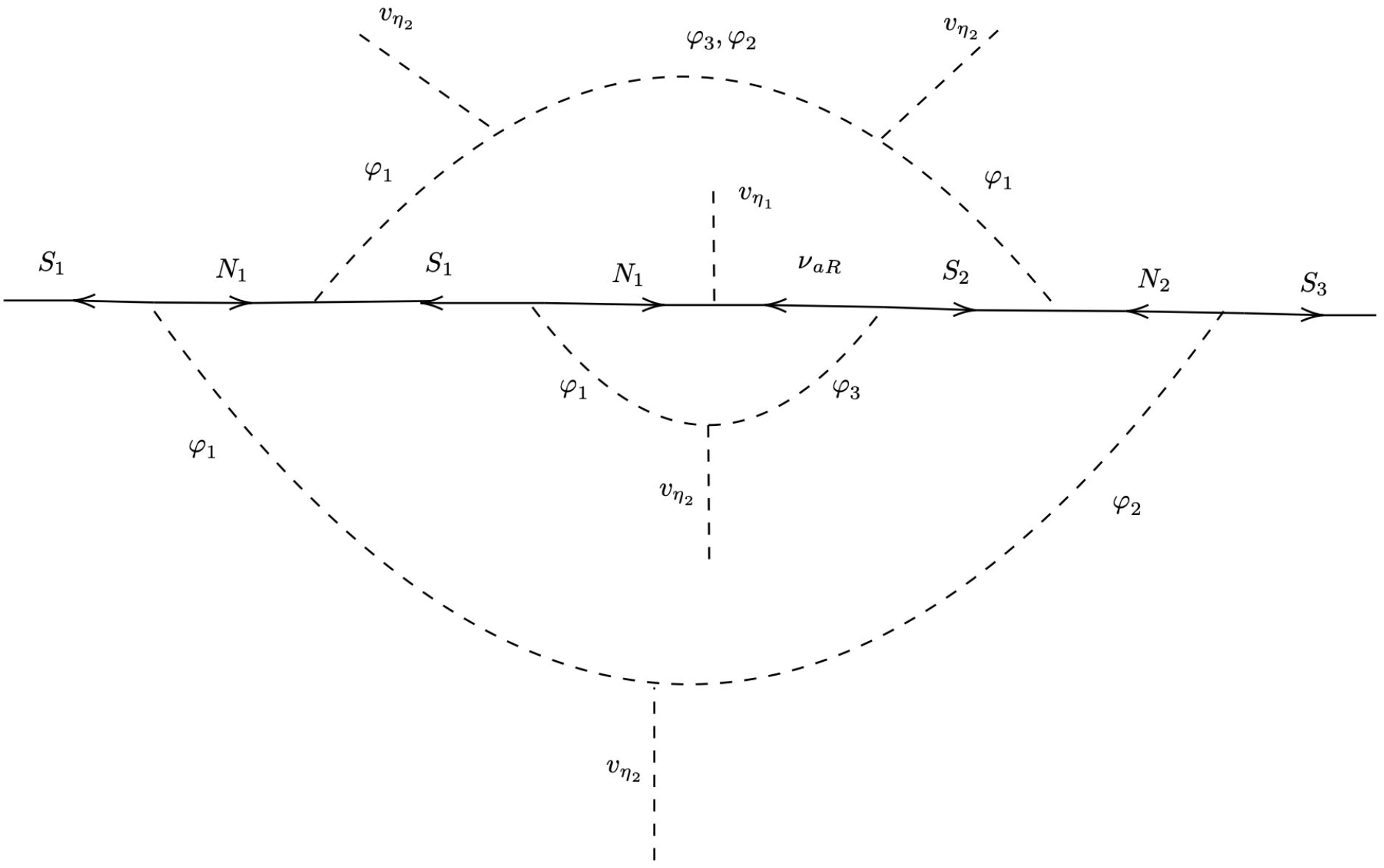}
			\includegraphics[scale=0.27]{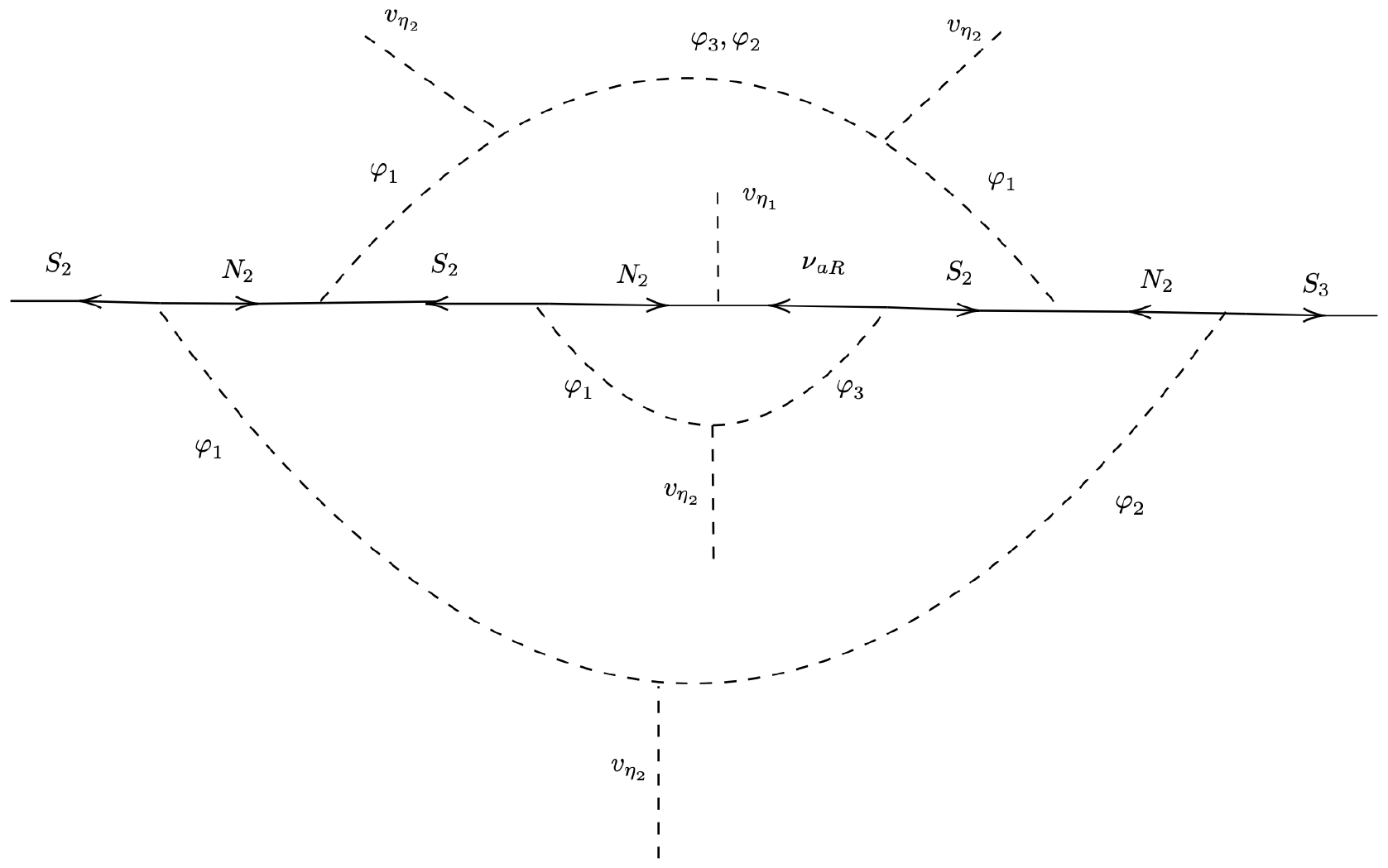}
	\end{center}
	\caption{Feynman diagrams contributing to the fermionic DM mass.}
	\label{DMmass}
\end{figure}
The $S_1^\prime, S_2^\prime,S_3^\prime $ are absolutely sterile in the sense that they do not mix with active neutrinos, and thus they decouple from SM particles, by conservation of the dark parity. 
Unlike typical sterile neutrinos, they are insensitive to all conventional constraints, a direct consequence of their mixing with active neutrinos \cite{Boyarsky:2009ix}. On the other hand, the $S_3^\prime$ is the lightest particle among those carrying odd dark parity, thus implying that $S^\prime_3$ is stable. With a mass at keV order, $S^\prime_3$ can be thermally generated in the early Universe due to $U(1)_N$ portal interactions mediated by a $Z^{\prime}$ exchange (see Appendix \ref{app3} for identification of $Z'$), which also brings $S^\prime_3$ in thermal equilibrium in the early Universe. The dark matter relic for the case of a relativistic freeze-out is computed by Kolb and Turner \cite{Kolb:1990vq}, and can be written as
\bea
\Omega_{DM}h^2 =76.4 \times \left( \frac{3g}{4g_{*s}(x_f)}\right)\left( \frac{m_{S_3^\prime}}{\text{keV}}\right),
\eea 
where $g$ is the number of internal degrees of freedom of the DM candidate, and $g_{*s}$ presents the number of relativistic entropy degrees of freedom at the epoch of $S_3^\prime$ decoupling, $x_f= \frac{m_{S_3}}{T_f}$, where $T_f$ is freeze-out temperature. If $S_3^\prime$ decouples around the electroweak scale, $g_{*s}\simeq 106.75$ and $m_{S_3^\prime}=5$ \text{keV}, the DM is overproduced by 100 times the correct DM abundance.  

\begin{figure}[h]
	\begin{center}
			\includegraphics[scale=0.6]{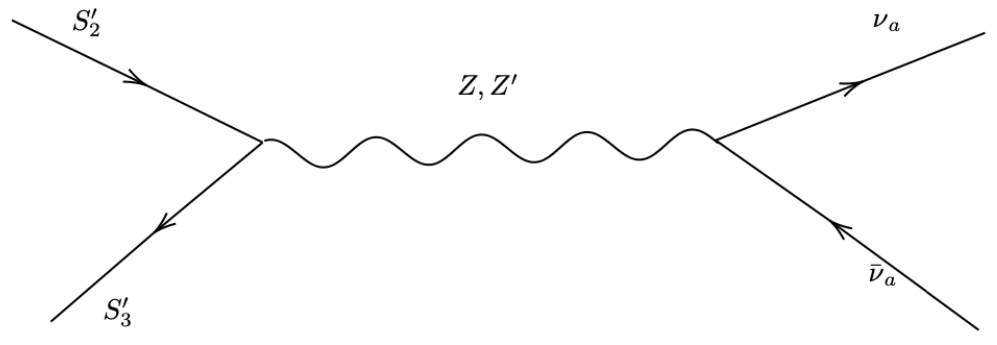}
			\includegraphics[scale=0.6]{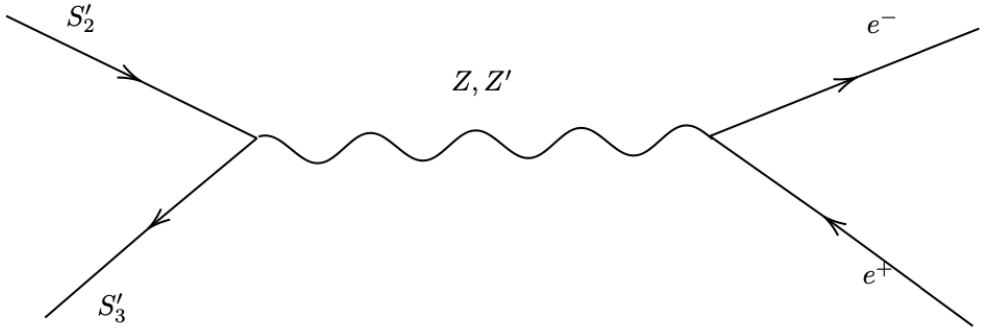}  
			\includegraphics[scale=0.7]{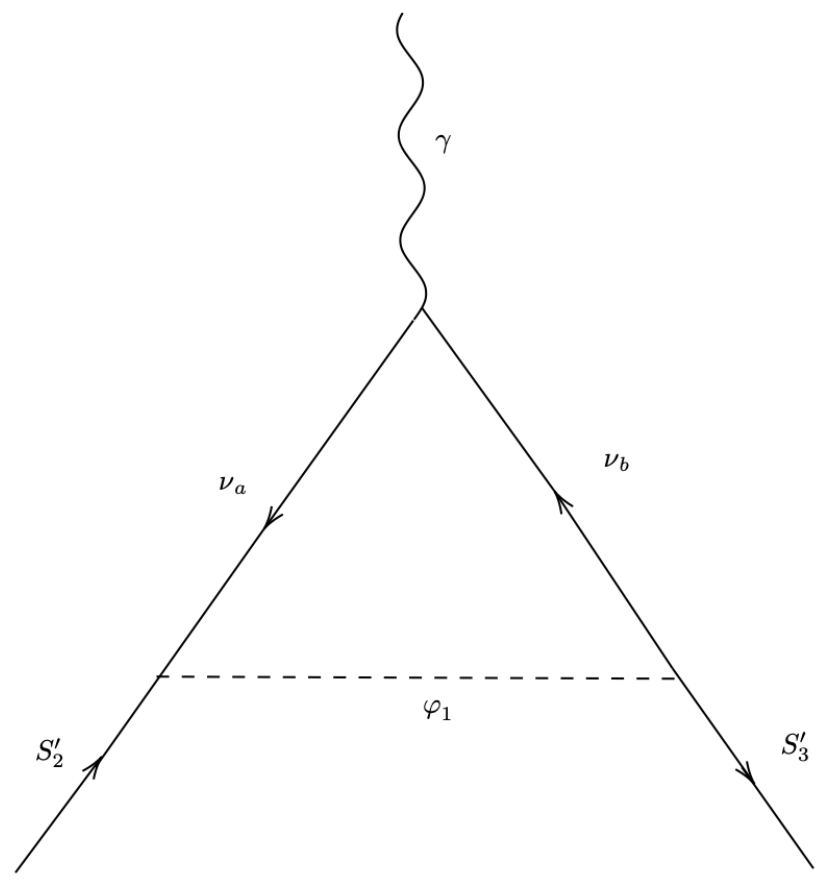}
	\end{center}
	\caption{Feynman diagrams describe the decay process of the long-lived particle.}
	\label{long-lived}
\end{figure}

This issue can be resolved through late-time entropy production resulting from the decay of long-lived heavier particles \cite{Borah:2021inn,Nemevsek:2012cd,Bezrukov:2009th,Dror:2020jzy,Dutra:2021lto,Arcadi:2021doo}. In our scenario, the field $S'_2$
naturally emerges as a long-lived particle. Its mass, generated radiatively at  the one-loop level, can range from MeV to GeV scales. Crucially, $S'_2$ is too light to decay into the neutral fermionic dark fields or the heavier scalar dark fields to which it couples, ensuring its longevity. 
Because of the mixing between $Z$ and $Z'$, and the mixing among the three neutral dark fermions $S_1, S_2, S_3$, the long-lived $S_2^\prime$ decays via the following modes: $S_2^\prime \to S_3^\prime \nu \bar{\nu}$; $S_2^\prime \to S_3^\prime \gamma$; and if $S_2^\prime$ is heavier than twice the electron mass, we also have the following decay channel: $S_{2}^\prime \to S_3^\prime e^+ e^-$, as depicted in Fig. \ref{long-lived}. The one-loop decay channel into a photon has a small branching ratio due to the suppressed couplings of $S_{2,3}\nu_a \varphi_1$; therefore, we will not consider its contribution.

Notice that such a decay should occur before the big bang nucleosynthesis, the temperature after the decay of long lived $S_2^\prime$ should be greater than few  $\text{MeV} $  in order to align with successful BNN predictions and can release extra entropy and dilute the abundance of $S^\prime_3$ to bring it into the observed limit \cite{Borah:2021inn,Nemevsek:2012cd,Bezrukov:2009th,Dror:2020jzy,Dutra:2021lto,Arcadi:2021doo}. 
It leads to the $S_2^\prime$ lifetime which should be shorter than two seconds. Let us consider the total decay width of the processes $S_2^\prime \to S_3^\prime \nu \nu$ and $S_{2}^\prime \to S_3^\prime e^+ e^-$. We obtain
\bea
\Gamma\left( S_2^\prime \to S_3^\prime e^+e^-\right) &=& \frac{G_F^2 m^5_{S_2^\prime}}{192 \pi^3}\left( \sin^2 \xi + \cos^2 \xi \frac{m_Z^4}{m_{Z^\prime}^4}\right)   \left| \sum _\delta  U_{2 \delta}^{S*}U_{3 \delta}^S\right|^2\nn \\  & \times& \left\{g_Lg_R I_2\left( 0, \xi_\al^2, \xi_\al^2\right) + (g_L^2+g_R^2) I_1\left( 0, \xi_\al^2, \xi_\al^2\right)\right\}, \\ 
\Gamma \left( S_2^\prime \to S_3^\prime \nu \nu \right) &=& \frac{G_F^2 m^5_{S_2^\prime}}{192 \pi^3}\left( \sin^2 \xi + \cos^2 \xi \frac{m_Z^4}{m_{Z^\prime}^4}\right)   \left|  \sum _\delta U_{2 \delta}^{S*}U_{3 \delta}^S\right|^2,
\eea
where $g_L = (-\frac{1}{2} + \sin^2 \theta_W) \cos \xi$, $g_R = \sin^2 \theta_W \cos \xi$, and the functions $I_{1,2} \left (x,y,z \right )$ can be found in \cite{Ballett:2019bgd}. We would like to note that to be consistent with the SM prediction on the $Z$-boson decay, the mixing angle $\xi$ must satisfy $\sin \xi < 10^{-2}$. Because the mass mixing matrix of neutral fermion dark fields has the form given in Equation (\ref{massDM}), the off-diagonal elements $U^S_{ij} \simeq 10^{-3}$ for $i \neq j$. To ensure the $S_2^\prime$ lifetime is shorter than two seconds, we obtain the constraint that the mass of $S_2^\prime$ should be larger than one GeV. This constraint on the long-lived mass satisfies the condition that the long-lived decay can release enough entropy to dilute the keV dark matter abundance.

If we assume that the real part of one of the scalar fields, $\varphi_i$, serves as the inflaton field and the potential governing the inflation involves non-minimal coupling between the inflaton and gravity, cosmic inflation is predicted similarly to that in the extended $U(1)_{B-L}$ symmetry model \cite{Borah:2021inn}. Therefore, if we impose inflationary conditions on the model parameters, the benchmark point to explain the dynamics of the phenomenology of DM is determined as previously done in \cite{Borah:2021inn}, specifically
\bea
g_N= 0.01, \hs m_{Z'} = 100\ \text{TeV}, \hs m_{S_3}= 5\ \text{keV}, \hs \Ga_{S_2}=10^{-22}\ \text{GeV}
\eea
with $g_N$ and $Z'$ to be the gauge coupling and boson of the $U(1)_N$ group, respectively.

\section{\label{concl} Conclusions} 

We have reconsidered a simple extension of the SM, called flipped standard model, by introducing an additional $U(1)_N$ gauge group, which is spontaneously broken down to a dark parity $Z_2$, i.e. $D_P=(-1)^{\fr 1 2 (5+3n_1)D+2s}$, where the dark charge $D$ is defined by $N$ through $T_3$, in similarity to the electric charge in the SM. 

The successful implementation of the radiative inverse seesaw mechanism to yield tiny active neutrino masses requires the inclusion of the neutral leptons $\nu_{aR}$, $N_n$ ($n=1,2$) and $S_a$  ($a=1,2,3$), with non trivial charges under the local $U(1)_N$ symmetry. These fields are also crucial for the cancellation of chiral anonamlies.

The considered model yields a fermionic dark matter candidate with very small mass at the keV scale, arising at three-loop level. The dark parity conservation prevents mixings of the keV DM candidate with active neutrinos as well as with parity even heavy neutrinos, then allowing 
to evade the cosmology and astrophysics bounds for sterile neutrinos. Due to the gauge interaction of DM, the DM may be thermally overproduced in the early universe. However, the later entropy dilution from the decay of long-lived particle can set the correct DM abundance. It is noted that the model can also address the issue of cosmic inflation, besides the keV DM phenomena.  

\section*{Acknowledgments}

This work is funded by the Vietnam Academy of Science and Technology, under Grant No. CBCLCA.03/25-27.
AECH is supported by ANID-Chile FONDECYT 1261103, FONDECYT 1241855, ANID---Millennium Science Initiative Program $ICN2019\_044$, ANID CCTVal CIA250027 and ICTP through the Associates Programme (2026-2031).

\appendix

\section{\label{app1}Lepton number violating Majorana mass submatrix $\mu$}
The lepton number violating (LNV) $2\times 2$ Majorana mass submatrix $\mu$ in our model only arises radiatively. Its $(1,1)$ entry is generated at one loop level, whereas the remaining entries arise at two loops. The entries of the LNV Majorana mass submatrix $\mu$ are given by
\begin{eqnarray}
\mu _{11}&=&\frac{\left( f_{1}^{N}\right) ^{2}m_{S_{1}}}{16\pi ^{2}}%
\sum\limits_{i=1}^{3}\left[ \left( \left( R_{H}\right) _{1i}\right) ^{2}%
\frac{m_{H_{i}}^{2}}{m_{H_{i}}^{2}-m_{S}^{2}}\ln \left( \frac{m_{H_{i}}^{2}}{%
	m_{S}^{2}}\right) -\left( \left( R_{A}\right) _{1i}\right) ^{2}\frac{%
	m_{A_{i}}^{2}}{m_{A_{i}}^{2}-m_{S}^{2}}\ln \left( \frac{m_{A_{i}}^{2}}{%
	m_{S}^{2}}\right) \right],\notag\\
&&
\end{eqnarray}
\begin{eqnarray}
\mu _{22} &=&\sum_{a=1}^{3}\sum_{s=1}^{3}\sum_{p=1}^{3}\frac{\left(
	f_{2}^{N}\right) ^{3}f_{2a}^{S}\left( M_{N}\right) _{a2}}{4(4\pi )^{4}} 
\notag \\
&&\times \int_{0}^{1}d\alpha \int_{0}^{1-\alpha }d\beta \frac{1}{\alpha
	(1-\alpha )}\Biggl[\left( \left( R_{H}\right) _{1s}\right) ^{2}\left(
R_{H}\right) _{1p}\left( R_{H}\right) _{3p}I\left( \left( \left(
M_{N}\right) _{a2}\right) ^{2},m_{R_{s}R_{p}}^{2},m_{R_{s}I_{p}}^{2}\right) 
\notag \\
&&-\left( \left( R_{A}\right) _{1s}\right) ^{2}\left( R_{A}\right)
_{1p}\left( R_{A}\right) _{3p}I\left( \left( \left( M_{N}\right)
_{a2}\right) ^{2},m_{I_{s}R_{p}}^{2},m_{I_{s}I_{p}}^{2}\right) \Biggr],
\end{eqnarray}
\begin{eqnarray}
\mu _{12} &=&\mu _{21}=\sum_{a=1}^{3}\sum_{s=1}^{3}\sum_{p=1}^{3}\frac{%
	\left( f_{1}^{N}\right) ^{2}f_{2}^{N}f_{2a}^{S}\left( M_{N}\right) _{a1}}{%
	4(4\pi )^{4}}  \notag \\
&&\times \int_{0}^{1}d\alpha \int_{0}^{1-\alpha }d\beta \frac{1}{\alpha
	(1-\alpha )}\Biggl[\left( \left( R_{H}\right) _{1s}\right) ^{2}\left(
R_{H}\right) _{1p}\left( R_{H}\right) _{3p}I\left( \left( \left(
M_{N}\right) _{a2}\right) ^{2},m_{R_{s}R_{p}}^{2},m_{R_{s}I_{p}}^{2}\right) 
\notag \\
&&-\left( \left( R_{A}\right) _{1s}\right) ^{2}\left( R_{A}\right)
_{1p}\left( R_{A}\right) _{3p}I\left( \left( \left( M_{N}\right)
_{a2}\right) ^{2},m_{I_{s}R_{p}}^{2},m_{I_{s}I_{p}}^{2}\right) \Biggr],
\end{eqnarray}
where the two-loop integral takes the form \cite{Kajiyama:2013rla,Hernandez:2021xet}, 
\begin{eqnarray}
I(m_{1}^{2},m_{2}^{2},m_{3}^{2}) 
&=& \frac{
m_{1}^{2}m_{2}^{2}\,\ln\!\left(\tfrac{m_{2}^{2}}{m_{1}^{2}}\right)
+ m_{2}^{2}m_{3}^{2}\,\ln\!\left(\tfrac{m_{3}^{2}}{m_{2}^{2}}\right)
+ m_{3}^{2}m_{1}^{2}\,\ln\!\left(\tfrac{m_{1}^{2}}{m_{3}^{2}}\right)
}{
(m_{1}^{2}-m_{2}^{2})(m_{1}^{2}-m_{3}^{2})
}\ , \notag \\[2mm]
m_{a_{k}b_{l}}^{2} 
&=& \frac{\beta\, m_{(S_{k})_{a}}^{2} + \alpha\, m_{(S_{l})_{b}}^{2}}{\alpha (1-\alpha)},
\qquad k,l=1,2,\quad a,b \in \{R,I\}\ .
\end{eqnarray}
and $H_{i}$ and $A_{i}$ ($i=1,2,3$) are the physical dark CP even and dark CP odd scalars, respectively. They are related with the real $\varphi_{iR}$ and imaginary $\varphi_{iI}$ components of the inert scalar singlets $\varphi_i$ ($i=1,2,3$) via the following relations, 
\begin{equation}
\left( 
\begin{array}{c}
\varphi _{1R} \\ 
\varphi _{2R} \\ 
\varphi _{3R}%
\end{array}%
\right) =R_{H}\left( 
\begin{array}{c}
H_{1} \\ 
H_{2} \\ 
H_{3}%
\end{array}%
\right) ,\hspace{1cm}\hspace{1cm}\left( 
\begin{array}{c}
\varphi _{1I} \\ 
\varphi _{2I} \\ 
\varphi _{3I}%
\end{array}%
\right) =R_{A}\left( 
\begin{array}{c}
A_{1} \\ 
A_{2} \\ 
A_{3}%
\end{array}%
\right) .
\end{equation}

\section{\label{app2} Mass matrix for dark neutral leptons}
Only the $(1,1)$ entry of the mass matrix for dark neutral leptons appearing in Eq. (\ref{massDM}) is generated at tree level, whereas the remaining ones arise radiatively. The entries of the dark neutral leptons mass matrix are given by
\begin{eqnarray}
m_{S_{1}S_{1}}^{tree} &=&f_1^S\frac{v_{\eta_2}}{\sqrt{2}},
\end{eqnarray}
\begin{eqnarray}
	m_{S_{2}S_{2}}^{1loop} &=&\sum\limits_{a=1}^{3}\frac{f_{2}^{N}f_{2a}^{S}%
		\left( M_{N}\right) _{a2}}{16\pi ^{2}}\left[ \sum\limits_{i=1}^{3}\left(
	R_{H}\right) _{1i}\left( R_{H}\right) _{3i}\frac{m_{H_{i}}^{2}}{%
		m_{H_{i}}^{2}-\left( \left( M_{N}\right) _{a1}\right) ^{2}}\ln \left( \frac{%
		m_{H_{i}}^{2}}{\left( \left( M_{N}\right) _{a1}\right) ^{2}}\right) \right. 
	\notag\\
	&&-\left. \left( R_{A}\right) _{1i}\left( R_{A}\right) _{3i}\frac{%
		m_{A_{i}}^{2}}{m_{A_{i}}^{2}-\left( \left( M_{N}\right) _{a1}\right) ^{2}}%
	\ln \left( \frac{m_{A_{i}}^{2}}{\left( \left( M_{N}\right) _{a1}\right) ^{2}}%
	\right) \right], 
\end{eqnarray}
\begin{eqnarray}
	m_{S_{1}S_{2}}^{1loop} &=&m_{S_{2}S_{1}}^{1loop} \\
	&=&\sum\limits_{a=1}^{3}\frac{f_{1}^{N}f_{2a}^{S}\left( M_{N}\right) _{a1}}{%
		16\pi ^{2}}\sum\limits_{i=1}^{3}\left[ \left( R_{H}\right) _{1i}\left(
	R_{H}\right) _{3i}\frac{m_{H_{i}}^{2}}{m_{H_{i}}^{2}-\left( \left(
		M_{N}\right) _{a2}\right) ^{2}}\ln \left( \frac{m_{H_{i}}^{2}}{\left( \left(
		M_{N}\right) _{a2}\right) ^{2}}\right) \right.\notag  \\
	&&-\left. \left( R_{A}\right) _{1i}\left( R_{A}\right) _{3i}\frac{%
		m_{A_{i}}^{2}}{m_{A_{i}}^{2}-\left( \left( M_{N}\right) _{a2}\right) ^{2}}%
	\ln \left( \frac{m_{A_{i}}^{2}}{\left( \left( M_{N}\right) _{a2}\right) ^{2}}%
	\right) \right], 
\end{eqnarray}
\begin{equation}
m_{S_{1}S_{3}}^{3loop}=m_{S_{3}S_{1}}^{3loop}\simeq \frac{\left(
	f_{1}^{N}\right) ^{3}f_{2}^{N}f_{3}^{S}}{\left( 16\pi ^{2}\right)
	^{3}}\sum_{a=1}^{3}f_{2a}^{S}\left( M_{N}\right) _{a1},
\end{equation}
\begin{equation}
m_{S_{2}S_{3}}^{3loop}=m_{S_{3}S_{2}}^{3loop}\simeq \frac{\left(
	f_{2}^{N}\right) ^{4}f_{3}^{S}}{\left( 16\pi ^{2}\right) ^{3}}%
\sum_{a=1}^{3}f_{2a}^{S}\left( M_{N}\right) _{a2},
\end{equation}
\begin{equation}
m_{S_{3}S_{3}}^{3loop}\simeq \frac{\left( f_{2}^{N}\right) ^{3}\left(
	f_{3}^{S}\right) ^{2}}{\left( 16\pi ^{2}\right) ^{3}}\sum_{a=1}^{3}f_{2a}^{S}\left(
M_{N}\right) _{a2}.
\end{equation}
Considering a simplified benchmark scenario where
\begin{equation}
f_{1}^N<<f_{2}^N\simeq f_{2a}^S<<f_{3}^S,\hspace{1cm}a=1,2,3,
\end{equation}
the mixing elements in the mass matrix for dark neutral leptons will be very small, then implying that the eigenvalues of the dark neutral lepton mass matrix can be approximated by its diagonal entries. Consequently, in this benchmark scenario the masses of the physical $S_1^\prime$, $S_2^\prime$ and $S_3^\prime$ dark neutral leptons can be estimated as follows
\begin{equation}
m_{S_1^\prime}\simeq  f_1^S\frac{v_{\eta_2}}{\sqrt{2}},
\end{equation}
\begin{eqnarray}
	m_{S_2^\prime}&\simeq&\sum\limits_{a=1}^{3}\frac{f_{2}^{N}f_{2a}^{S}%
		\left( M_{N}\right) _{a2}}{16\pi ^{2}}\left[ \sum\limits_{i=1}^{3}\left(
	R_{H}\right) _{1i}\left( R_{H}\right) _{3i}\frac{m_{H_{i}}^{2}}{%
		m_{H_{i}}^{2}-\left( \left( M_{N}\right) _{a1}\right) ^{2}}\ln \left( \frac{%
		m_{H_{i}}^{2}}{\left( \left( M_{N}\right) _{a1}\right) ^{2}}\right) \right. 
	\notag\\
	&&-\left. \left( R_{A}\right) _{1i}\left( R_{A}\right) _{3i}\frac{%
		m_{A_{i}}^{2}}{m_{A_{i}}^{2}-\left( \left( M_{N}\right) _{a1}\right) ^{2}}%
	\ln \left( \frac{m_{A_{i}}^{2}}{\left( \left( M_{N}\right) _{a1}\right) ^{2}}%
	\right) \right], 
\end{eqnarray}
\begin{equation}
m_{S_3^\prime}\simeq\frac{\left( f_{2}^{N}\right) ^{3}\left(
	f_{3}^{S}\right) ^{2}}{\left( 16\pi ^{2}\right) ^{3}}\sum_{a=1}^{3}f_{2a}^{S}\left(
M_{N}\right) _{a2}.
\end{equation}

\section{\label{app3} Diagonalizing neutral gauge bosons}

The model contains three neutral gauge components that are mixed via a $3 \times 3$ matrix. Diagonalizing this mixing neutral gauge mass matrix, we obtain the photon $A_\mu$ and two massive neutral gauge bosons, $Z_\mu$ and $Z^\prime_\mu$, which are related to the flavor states as follows
	\begin{equation*}
		\begin{pmatrix}
			A_3 \\ B \\ B_N
		\end{pmatrix}
		=
		\begin{pmatrix}
			\sin \theta_W & \cos \theta_W \cos \xi & -\cos \theta_W \sin \xi \\
			\cos \theta_W & -\sin \theta_W \cos \xi & \sin \theta_W \sin \xi \\
			0 & \sin \xi & \cos \xi
		\end{pmatrix}
		\begin{pmatrix}
			A \\ Z \\ Z'
		\end{pmatrix},
	\end{equation*}
	where $\tan 2 \xi =\frac{2 g g_N v^2 c_W}{g_N^2 c^2_W [v^2 + 16 \delta (\Lambda_1 + 2 \Lambda_2)] - g^2 v^2 }.$ Here, $g_N$ and $B_N$ are the gauge coupling and the gauge field associated with the $U(1)_N$ group, respectively. 
	
\vspace{-0.4cm} 
\bibliographystyle{JHEP}
\bibliography{combine}

\end{document}